\documentclass[sigconf]{acmart}

\AtBeginDocument{%
  }

\setcopyright{none}
\renewcommand\footnotetextcopyrightpermission[1]{}
\usepackage{multirow}
\usepackage{placeins}
\usepackage{float}

\renewcommand{\shortauthors}{Abdallah et al.}

\begin{document}

\title{The Hidden Environmental Cost of Poor Coding Practices in TensorFlow and Keras Applications: A Study on Resource Leaks and Carbon Emissions}

\author{Bashar Abdallah}
\affiliation{
  \institution{Department of Computer Engineering and Software Engineering, Polytechnique Montréal}
  \city{Montréal}
  \country{Canada}
}
\email{bashar.abdallah@etud.polymtl.ca}

\author{Gustavo Santos}
\affiliation{
  \institution{Department of Computer Engineering and Software Engineering, Polytechnique Montréal}
  \city{Montréal}
  \country{Canada}
}
\email{gustavo.palma-dos-santos@etud.polymtl.ca}

\author{Rola Al Bataineh}
\affiliation{
  \institution{Department of IT and Software Engineering, École de Technologie Supérieure (ETS), Université du Québec}
  \city{Montréal}
  \country{Canada}
}
\email{rola.al-bataineh.1@ens.etsmtl.ca}

\author{Alain Abran}
\affiliation{
  \institution{Department of IT and Software Engineering, École de Technologie Supérieure (ETS), Université du Québec}
  \city{Montréal}
  \country{Canada}
}
\email{alain.abran@etsmtl.ca}

\author{Mohammad Hamdaqa}
\affiliation{
  \institution{Department of Computer Engineering and Software Engineering, Polytechnique Montréal}
  \city{Montréal}
  \country{Canada}
}
\email{mhamdaqa@polymtl.ca}

\acmConference[EASE '26]{International Conference on Evaluation and Assessment in Software Engineering}{June 09--12, 2026}{Glasgow, United Kingdom}
\begin{abstract}
Efficiency and sustainability are critical considerations in the development and deployment of machine learning (ML) applications. Among the factors influencing sustainability, resource leaks in ML code can introduce hidden inefficiencies that elevate energy consumption and CO$_2$ emissions. Despite this, empirical evidence quantifying their environmental impact remains limited. This emerging results paper presents an initial empirical investigation of two common resource-leak smells (i.e., Improper Model Reuse (IMR) and Unreleased Tensor References (UTR)) and their impact on energy consumption and CO$_2$ emissions in TensorFlow and Keras workloads. Controlled experiments were conducted for each smell by executing identical training tasks while comparing against a smell-free baseline. Our preliminary results show that both smells consistently increase estimated electricity usage and carbon emissions. IMR and UTR increased electricity consumption by approximately 32\% and 46\%, respectively, with proportional increases in CO$_2$ emissions. Paired statistical tests indicate that these differences are systematic and statistically significant, providing initial empirical evidence that resource-leak smells may degrade ML energy efficiency and environmental sustainability. These findings suggest that resource-leak smells pose measurable risks to both software quality and sustainability, emphasizing the importance of integrating resource-lifecycle management and energy-efficiency considerations into ML development.
\end{abstract}

\maketitle

\begin{center}
\small\textit{Accepted at the International Conference on Evaluation and Assessment in Software Engineering (EASE 2026). This is the author version.}
\end{center}

\section{Introduction}

The rapid expansion of machine learning (ML) systems has intensified concerns regarding their environmental footprint. Training large-scale ML models can consume substantial amounts of electricity; often in the order of 1{,}000--2{,}000~MWh—and generate significant CO\(_2\) emissions~\cite{strubell2020,patterson2021}. While prior research has primarily focused on hardware efficiency and algorithmic optimization, considerably less attention has been given to how software engineering practices influence the sustainability of ML systems.

Recent work~\cite{anonymous2025} identified a class of ML-specific code smells that lead to resource leaks in frameworks such as PyTorch, TensorFlow, and Keras. These smells can cause unintended GPU retention and excessive tensor accumulation, suggesting potential increases in computational overhead. However, the environmental implications of such resource-leak smells remain unquantified. In particular, it is unclear whether these coding practices produce measurable increases in energy consumption and CO\(_2\) emissions under controlled execution conditions.

To address this gap, we adopt a controlled within-subject (repeated-measures) design, which is recommended for isolating treatment effects in software engineering experiments~\cite{wohlin2012experimentation}. We evaluate two ML-specific resource-leak smells in TensorFlow/Keras applications: Improper Model Reuse (IMR) and Unreleased Tensor References (UTR). Each smell-injected configuration is systematically compared against a functionally equivalent baseline while monitoring electricity consumption and CO$_2$ emissions during execution. Our analysis estimates both the magnitude and statistical significance of the observed sustainability differences, enabling a controlled empirical assessment of the environmental overhead introduced by resource mismanagement.

This paper makes the following contributions:

\begin{itemize}
    \item We provide a controlled empirical quantification of the energy and CO$_2$ impact of ML-specific resource-leak smells.
    \item We demonstrate, under controlled experimental conditions, that resource lifecycle mismanagement can introduce substantial sustainability overhead without statistically significant improvements in predictive performance.
    \item We highlight the importance of integrating sustainability-oriented evaluation into ML software engineering practices.
\end{itemize}

Overall, our findings provide controlled preliminary evidence that seemingly minor coding decisions in ML systems can translate into measurable environmental costs, reinforcing the need to treat resource management as a first-class software quality concern.

\section{Related Work}
Code smells are widely recognized as indicators of poor software quality and long-term maintainability risks~\cite{suryanarayana2014}. A broad body of work surveys traditional code smells and ML-based detection techniques, such as the systematic review in~\cite{yadav2024}, which highlights common smells (e.g., God Class, Long Method) and the growing use of learning-based classifiers for their identification. Similarly, several studies have examined code quality issues in machine learning contexts. The authors in~\cite{jebnoun2020} identified deep-learning-specific smells, including hard-coded hyperparameters and monolithic scripts, showing that performance-driven development often compromises software engineering practices. More recently, \cite{anonymous2025} identified resource-leak smells in TensorFlow, Keras, and PyTorch that can lead to unintended GPU retention and excessive tensor accumulation. Their work established the existence of these ML-specific resource-leak smells but did not quantify their energy or carbon impact, leaving a gap that motivates further investigation.

Beyond ML frameworks, research has shown that some code smells can influence resource consumption. For example, \cite{misu2025} demonstrated that smelly test cases consume more energy than non-smelly ones, reinforcing the link between poor coding practices and inefficient resource usage. Complementary lines of work have investigated the environmental footprint of ML systems. Studies such as~\cite{sanchez2025,mavromatis2024} reveal that ML workflows can incur substantial energy cost, yet evaluation pipelines often overlook sustainability metrics.

Overall, while existing research addresses (1) code smell detection, (2) resource overuse caused by some code smells in software, and (3) sustainability concerns in ML operations, no prior work has quantified the energy and CO$_2$ implications of ML-specific resource-leak smells. 

\section{Methodology}

This study aims to empirically evaluate whether resource-leak smells in TensorFlow/Keras applications increase energy consumption and CO$_2$ emissions under controlled conditions. To achieve this goal, we conducted a repeated-measures experiment comparing a clean baseline model against two independently injected resource-leak smells previously identified in~\cite{anonymous2025}. All configurations were executed under identical hardware, dataset, and initialization conditions, enabling direct paired comparisons of sustainability metrics.

We followed a structured experimental procedure. First, we implemented a clean baseline model and established reference sustainability measurements under controlled execution conditions. Second, each resource-leak smell was injected independently while preserving architectural and training equivalence. Third, all configurations were executed ten times in independent processes using fixed random seeds to ensure reproducibility and mitigate stochastic variation. During each run, electricity consumption and CO$_2$ emissions were monitored using CodeCarbon. Finally, paired statistical analyses were performed to quantify the magnitude and statistical significance of the sustainability differences between baseline and smell-injected configurations. Figure~1 provides an overview of the research methodology.

\begin{figure}[t]
    \centering
    \includegraphics[width=.3\textwidth]{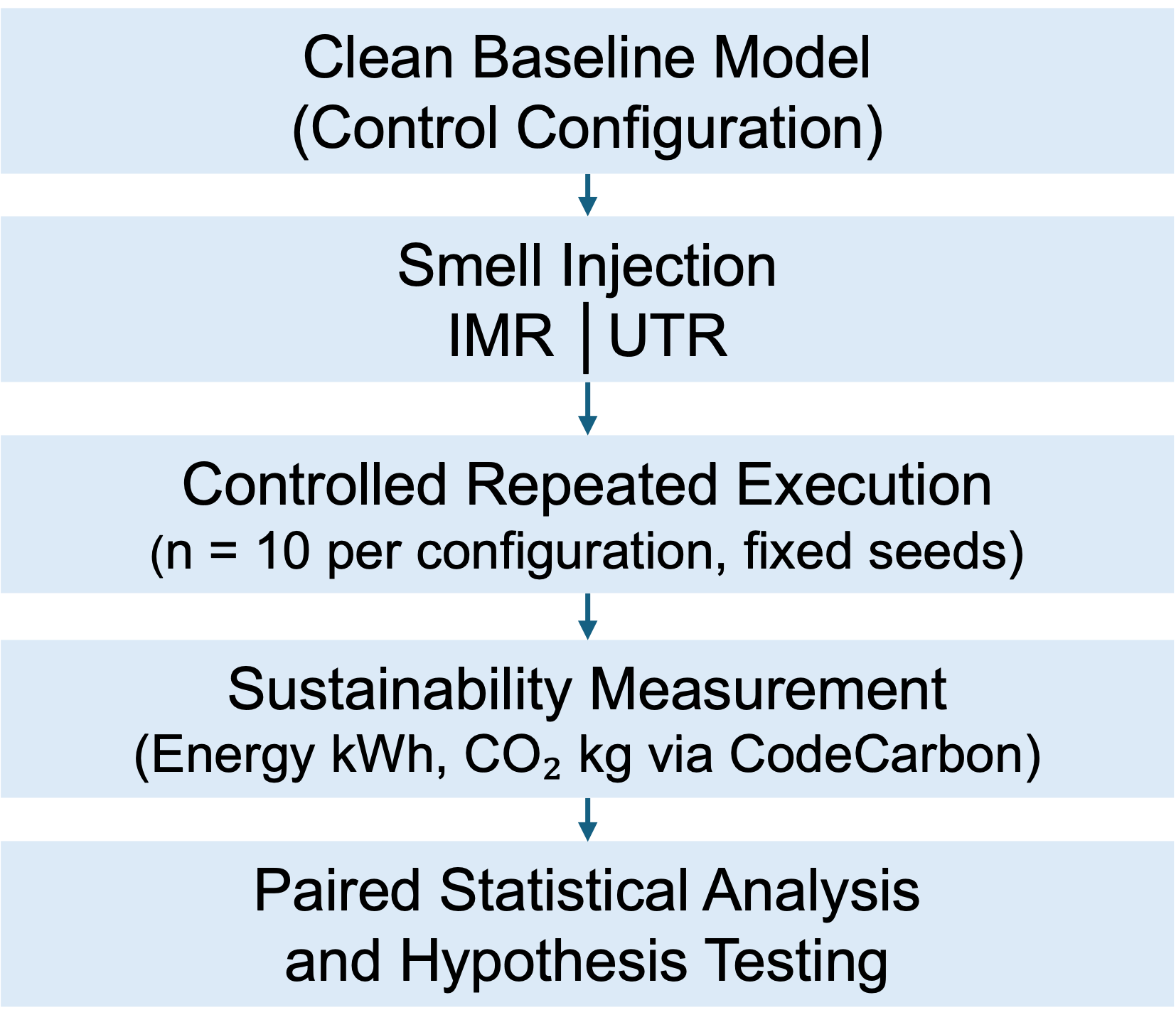}
    \caption{Overview of the research methodology.}
\end{figure}

\subsection{Experimental Design}

We adopted a controlled within-subject (repeated-measures) design to isolate the sustainability impact of each resource-leak smell. For every run index $i$, each smell-injected configuration (IMR and UTR) was paired with its corresponding baseline execution under identical hardware, dataset, software environment, and random seed conditions. This pairing controls for stochastic variation across runs and enables precise estimation of sustainability differences attributable to the injected smell.

Each configuration was executed ten times in independent processes to mitigate runtime variability and to support statistical inference. Random seeds for NumPy, TensorFlow, and dataset splits were fixed to preserve functional equivalence across runs. Minor nondeterminism (e.g., GPU scheduling or garbage-collection timing) may still affect runtime measurements; however, the repeated-measures design reduces its influence on the paired comparisons.

For each run, electricity consumption (kWh) and CO$_2$ emissions (kg) were recorded. Paired differences were computed as:

\[
d_i = X_{i,\text{smell}} - X_{i,\text{baseline}},
\]

where $X$ represents the sustainability metric of interest. We report the mean paired difference $\bar{d}$, relative percentage change over the baseline, and 95\% confidence intervals.

Directional hypotheses were evaluated using one-sided paired-sample $t$-tests with $df = n - 1$. Assumptions of approximate normality of paired differences were examined, and Wilcoxon signed-rank tests were additionally performed to assess robustness to distributional deviations. To quantify standardized effect magnitude, Cohen’s $d_z = \bar{d} / s_d$ was computed for each comparison.

\subsection{Smell Selection Criteria}

Given the short-paper format and planned future extensions, code-smell selection was deliberately limited to smells satisfying the criteria in Table~\ref{tab:criteria}: (C1) injectable without disrupting execution, (C2) inducing measurable computational overhead, (C3) independent of other smells, and (C4) compatible with TensorFlow~2.x eager execution. These criteria ensure that each smell can be evaluated in isolation without altering the intended functional behavior of the baseline model.

\begin{table}[h!]
\centering
\caption{Code smell inclusion criteria.}
\label{tab:criteria}
\begin{tabular}{@{}p{0.8cm}p{2.4cm}p{4.5cm}@{}}
\toprule
No. & Criterion & Description / Rationale \\
\midrule
C1 & Injectability & Can be introduced in TensorFlow~2.x without execution failure. \\
\addlinespace
C2 & Measurable Overhead & Produces $\geq 5\%$ change in runtime or CO$_2$ emissions. \\
\addlinespace
C3 & Isolation & Does not interact with other smells. \\
\addlinespace
C4 & TF~2.x Compatibility & Uses eager execution; no TF~1.x dependencies. \\
\bottomrule
\end{tabular}
\end{table}

\subsection{Baseline Configuration}

The baseline configuration consisted of a TensorFlow~2.x/Keras convolutional neural network (CNN) trained on the CIFAR-10 dataset, which contains 60{,}000 color images (32$\times$32 pixels) across 10 classes, with 50{,}000 images for training and 10{,}000 for testing. Images were normalized to the range $[0,1]$, labels were one-hot encoded, and on-the-fly data augmentation (random horizontal flips and spatial shifts) was applied during training to improve generalization.

The CNN architecture comprised three convolutional blocks with increasing filter sizes (32, 64, and 128). Each block included two convolutional layers with ReLU activation, followed by batch normalization, max-pooling, and dropout (rate = 0.25). The convolutional backbone was followed by a fully connected layer with 128 ReLU-activated units, a dropout layer (rate = 0.25), and a final Dense layer with 10 units and softmax activation for multi-class classification.

The model was trained for a total of 50 epochs using a batch size of 32, the Adam optimizer, and categorical cross-entropy loss. To ensure structural equivalence across configurations—particularly with the IMR treatment—the 50 training epochs were executed as five sequential cycles of 10 epochs each. This design choice ensures that differences in energy consumption are attributable to resource management behavior rather than differences in training control flow.

Each experimental run was executed as an independent process. Random seeds for NumPy, TensorFlow, and dataset splits were fixed for all runs to preserve functional equivalence and identical weight initialization across configurations.

\subsection{Resource-Leak Treatments}

Two resource-leak smells previously identified in~\cite{anonymous2025} were injected independently into the baseline configuration. Each smell was introduced in isolation, and no configuration combined multiple smells. Apart from the injected modification, all architectural, hyperparameter, dataset, and execution settings were identical to the baseline.

\textbf{Improper Model Reuse (IMR):}
The IMR smell was operationalized by reusing the same \texttt{tf.keras} model instance across the five sequential training cycles (10 epochs each) without reinitializing weights, recompiling the model, or clearing the underlying Keras session between cycles. This design mimics improper lifecycle management in which model objects persist across repeated training invocations, potentially leading to accumulation of computational graph artifacts or intermediate state.

To preserve functional equivalence, each experimental run was executed as an independent operating-system process with identical random seed initialization. No cumulative learning occurred across runs. The optimizer configuration, loss function, and data pipeline were unchanged relative to the baseline. Consequently, IMR alters resource lifecycle management behavior without modifying the intended optimization trajectory or training semantics.

\textbf{Unreleased Tensor Reference (UTR):}
The UTR smell was implemented via a custom \texttt{tf.keras.callbacks.Callback} that intercepted training at the batch level using the \texttt{on\_train\_batch\_end} hook. Every five batches, a manual forward pass was executed on the current mini-batch, and the resulting prediction tensors were appended to a persistent Python list maintained for the duration of the run. By maintaining active references to these tensors, TensorFlow's memory management mechanism was prevented from reclaiming the associated resources.

The callback did not modify gradients, optimizer state, loss computation, or model parameters. Its sole purpose was to retain intermediate tensor references, thereby simulating realistic misuse patterns that lead to unintended memory retention and increased resource consumption.

All treatment executions preserved identical dataset splits, batch sizes, augmentation procedures, and training duration as the baseline.

\subsection{Experimental Environment}

All experiments were conducted on a dedicated machine equipped with an NVIDIA Tesla T4 GPU. No concurrent user processes or GPU-intensive workloads were executed during experimentation to minimize measurement noise and external interference. The software environment consisted of TensorFlow~2.x, Keras, NumPy, Pandas, and Matplotlib, running under a consistent Python environment across all configurations.

Environmental metrics were collected using CodeCarbon~\cite{codecarbon2024}. CodeCarbon estimates electricity consumption (kWh) by sampling hardware utilization and power draw through NVIDIA Management Library (NVML) interfaces and applying hardware-specific power characteristics. CO$_2$ emissions are computed as a deterministic linear transformation of measured energy consumption using a fixed regional carbon intensity factor. Because this transformation is linear, inferential statistics for CO$_2$ emissions are mathematically equivalent to those for energy consumption; both metrics are reported for completeness and interpretability.

Each experimental run was executed as an independent operating-system process to prevent residual GPU state or memory artifacts from influencing subsequent runs. Before each run, GPU memory was cleared and the environment reinitialized. Random seeds for NumPy and TensorFlow were fixed to ensure identical initialization across configurations.

\subsection{Hypotheses}

To evaluate the sustainability impact of resource-leak smells, we formulated directional hypotheses for both electricity consumption and CO$_2$ emissions.

\textbf{Energy Consumption}

\textbf{H1:} Models containing resource-leak smells consume significantly more electricity during execution than equivalent smell-free baseline models.

\textbf{H0$_1$:} There is no significant difference in electricity consumption between smell-injected and baseline models.

\textbf{CO$_2$ Emissions}

\textbf{H2:} Models containing resource-leak smells produce significantly higher CO$_2$ emissions during execution than equivalent smell-free baseline models.

\textbf{H0$_2$:} There is no significant difference in CO$_2$ emissions between smell-injected and baseline models.

Because CO$_2$ emissions are computed as a deterministic linear transformation of energy consumption under a fixed regional carbon intensity factor, statistical outcomes for both metrics are expected to exhibit identical inferential patterns. Both metrics are reported for completeness and interpretability.

\subsection{Statistical Analysis}

The experiment followed a paired repeated-measures design. For each run index $i$ ($n = 10$), paired differences were computed as:

\[
d_i = X_{i,\text{smell}} - X_{i,\text{baseline}},
\]

where $X$ represents the sustainability metric of interest (energy in kWh or CO$_2$ in kg).

For each smell and metric, we computed:

- The mean paired difference $\bar{d}$
- The standard deviation of paired differences $s_d$
- The standard error $SE_{\bar{d}} = s_d / \sqrt{n}$
- The relative percentage change over baseline
- The 95\% confidence interval (CI) for $\bar{d}$

This estimation-focused approach emphasizes the magnitude and precision of observed differences rather than relying solely on $p$-values.

Directional hypotheses were evaluated using one-sided paired-sample $t$-tests with degrees of freedom $df = n - 1$. Because CO$_2$ emissions are computed as a deterministic linear transformation of energy consumption under a fixed carbon intensity factor, inferential testing was conducted primarily on energy consumption, while CO$_2$ results are reported for interpretability and environmental contextualization. The normality assumption for paired differences was examined, and Wilcoxon signed-rank tests were additionally performed to assess robustness to potential distributional deviations.

To quantify standardized effect magnitude, Cohen’s paired-sample effect size was computed as:

\[
d_z = \frac{\bar{d}}{s_d},
\]

which expresses the mean paired difference relative to its variability. Effect sizes are reported alongside confidence intervals and hypothesis tests to provide complementary perspectives on practical and statistical significance.

\section{Results}

\label{sec:results}
This section presents the sustainability impact of the injected resource‑leak smells, focusing on how these smells influence energy consumption, carbon emissions, and execution efficiency. All reported estimations represent averages across ten independent runs to ensure statistical robustness and reduce variance across executions.

\subsection{Sustainability Impact}

\subsubsection{Descriptive Results}

Energy values are reported in kWh (and converted to Wh for interpretability). CO$_2$ emissions are reported in kg (converted to grams in text).

The baseline configuration consumed on average 0.025646~kWh (25.65~Wh) per run and emitted 0.011608~kg (11.61~g) of CO$_2$.

IMR increased mean energy consumption to 0.033797~kWh (33.80~Wh) and CO$_2$ emissions to 0.015297~kg (15.30~g), representing a 31.78\% increase over baseline.

UTR produced the largest increases, with mean energy consumption of 0.037393~kWh (37.39~Wh) and CO$_2$ emissions of 0.016921~kg (16.92~g), corresponding to a 45.80\% increase in energy and a 45.77\% increase in CO$_2$ relative to baseline.

Because CO$_2$ emissions are computed by CodeCarbon as a deterministic linear transformation of energy consumption under a fixed regional carbon intensity factor, proportional increases in energy directly translate into proportional increases in emissions. As a result, inferential statistics for CO$_2$ are mathematically equivalent to those for energy. We report both metrics for completeness and interpretability, but statistical significance patterns are expected to mirror one another.

\begin{figure}[h]
    \centering
    \includegraphics[width=.85\linewidth]{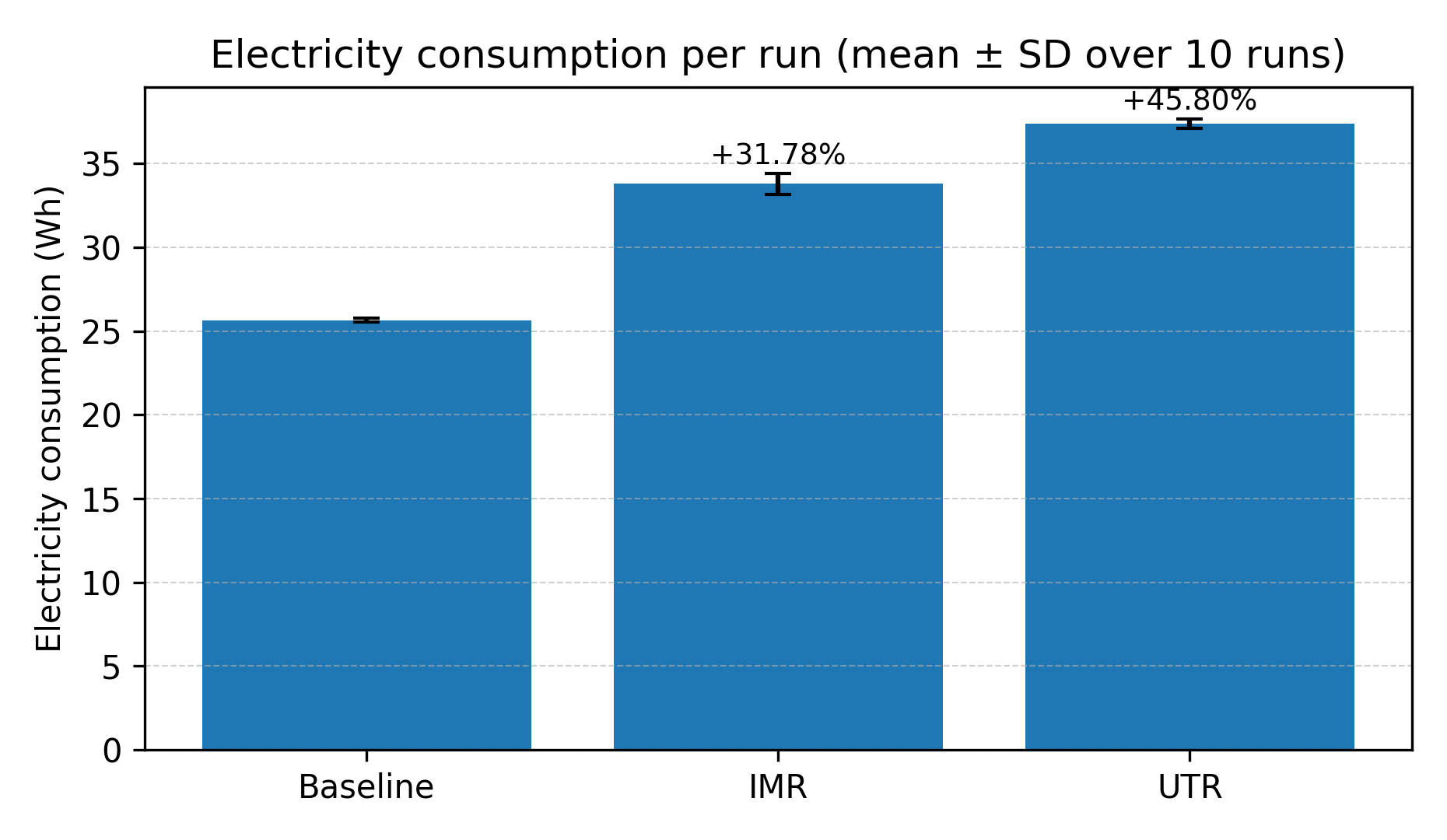}
    \caption{Electricity consumption per run (mean ± SD across ten runs). IMR increased energy by 31.78\% and UTR by 45.80\% relative to baseline.}
    \label{fig:electricity_N}
\end{figure}

\begin{figure}[h]
    \centering
    \includegraphics[width=.85\linewidth]{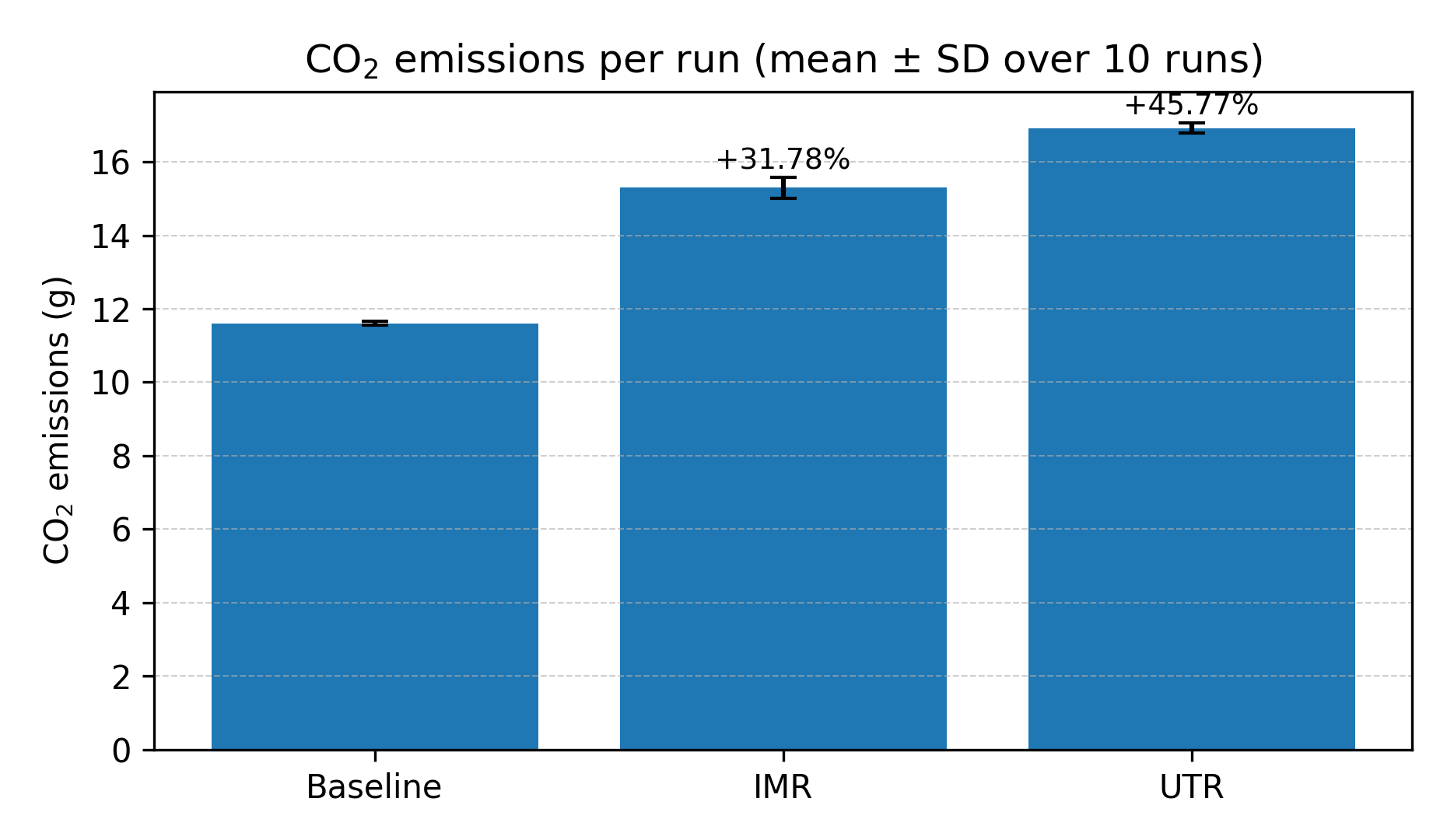}
    \caption{CO$_2$ emissions per run (mean ± SD across ten runs). IMR increased emissions by 31.78\% and UTR by 45.77\% relative to baseline.}
    \label{fig:co2_N}
\end{figure}

Figure~\ref{fig:electricity_N} illustrates electricity consumption across configurations, and Figure~\ref{fig:co2_N} presents the corresponding CO$_2$ emissions.

\subsubsection{Paired Statistical Analysis}

Paired differences were computed per run to control stochastic variation.

\begin{table}[t]
\centering
\caption{Paired Differences for Sustainability Metrics}
\label{tab:sustainability_stats}
\footnotesize
\begin{tabular*}{\columnwidth}{@{\extracolsep{\fill}}lcccc}
\hline
Comparison & Mean Diff. & SD & 95\% CI & $t(9)$ \\
\hline
IMR Energy (kWh) & 0.008150 & 0.000683 & [0.007663, 0.008638] & 37.83 \\
UTR Energy (kWh) & 0.011746 & 0.000358 & [0.011490, 0.012002] & 103.79 \\
IMR CO$_2$ (kg) & 0.003689 & 0.000309 & [0.003468, 0.003910] & 37.83 \\
UTR CO$_2$ (kg) & 0.005313 & 0.000158 & [0.005196, 0.005429] & 103.10 \\
\hline
\end{tabular*}
\end{table}

All paired $t$-tests were statistically significant ($p < 0.001$). The large $t$-statistics are attributable to the low variability of the paired differences across runs. Because all experiments were conducted under tightly controlled conditions—identical hardware, fixed random seeds, deterministic training cycles, and isolated GPU usage—the injected resource-leak smells introduced highly consistent additional energy overhead.

The small variance observed in the paired differences is therefore primarily a consequence of the paired experimental design and controlled execution environment. Each smell-injected run was directly matched with a baseline execution under identical conditions, which effectively cancels out most environmental and stochastic noise sources (e.g., scheduling jitter or minor nondeterminism). As a result, the remaining variability reflects systematic overhead introduced by the resource-leak treatments rather than uncontrolled runtime fluctuations.

All reported $t$-tests correspond to the one-sided hypotheses defined in Section~3.1. Wilcoxon signed-rank tests yielded consistent results ($p < 0.001$), confirming robustness to distributional assumptions.

Cohen's $d_z$ effect sizes were large for both smells (IMR: 11.93; UTR: 32.80 for energy), with nearly identical values for CO$_2$ due to the linear relationship between the two metrics. These magnitudes further reflect the stability and consistency of the injected overhead across runs.


\subsection{Summary of Findings}

The injected resource-leak smells produced substantial and statistically significant increases in energy consumption and CO$_2$ emissions (31.78–45.80\%). 

These findings indicate that resource-leak smells introduce considerable environmental overhead and increase the operational costs of ML applications.

Detailed per-run measurements and full statistical calculations are provided in Appendix~\ref{appendix:raw_results}.

\section{Discussion}

This section interprets the empirical findings in light of the study objectives. We discuss the sustainability impact of resource-leak smells and reflect on the implications for ML software engineering.

\subsection{Sustainability Impact of Resource-Leak Smells}

Consistent with the hypotheses defined in Section~3, this study provides controlled empirical evidence that resource-leak smells introduce measurable environmental overhead. Compared to the baseline configuration (25.65 Wh and 11.61 g CO$_2$ per run), IMR increased energy consumption and emissions by 31.78\%, while UTR increased them by approximately 46\%.

The paired analysis demonstrated that these increases were consistent across all ten runs, with narrow confidence intervals and statistically significant differences ($p < 0.001$). The magnitude and stability of the paired differences indicate that the observed overhead is systematic rather than attributable to stochastic runtime variation.

Because CO$_2$ emissions in CodeCarbon are computed as a linear function of energy consumption under fixed carbon intensity factors, proportional increases in energy directly translate into proportional increases in emissions. Consequently, the statistical patterns for both metrics are nearly identical. This proportionality reinforces the interpretation that the primary driver of increased environmental impact is computational inefficiency introduced by resource mismanagement.

Taken together, the results provide strong empirical support for H1 and H2, indicating that resource-leak smells significantly increase energy consumption and carbon emissions and thereby rejecting H0$_1$ and H0$_2$.

\subsection{Implications for ML Software Engineering}

The results extend the interpretation of resource-leak smells beyond maintainability and runtime correctness. Resource-leak smells demonstrably degrade computational efficiency, leading to 32–46\% increases in per-run energy consumption. In production or large-scale training settings, such proportional increases could accumulate into substantial operational costs and environmental impact.

From a software engineering perspective, resource management should therefore be treated as a first-class quality attribute alongside correctness and performance. Integrating smell detection tools, enforcing resource cleanup practices, and incorporating energy monitoring into development workflows may reduce hidden sustainability debt in ML systems.

Overall, the findings support the view that ML code quality is not only a maintainability concern but also an environmental responsibility.

\section{Threats to Validity}

\textbf{Construct Validity.}
Energy consumption and CO$_2$ emissions were estimated using CodeCarbon rather than measured with external power instrumentation. CodeCarbon infers electricity usage from hardware utilization samples and applies regional carbon intensity factors. While widely adopted in ML sustainability research, these values remain estimates. To improve reliability, experiments were conducted on an NVIDIA GPU with NVML-based monitoring, and runs were executed in isolation to minimize noise. Absolute values should therefore be interpreted cautiously, while relative comparisons remain robust.

\textit{Sensitivity to Measurement Noise.}
Although CodeCarbon estimates rather than directly measures power draw, our paired comparisons were conducted under identical hardware and execution conditions, so any systematic estimation bias would affect both configurations equally. Even under a conservative hypothetical measurement error of $\pm5\%$, the observed increases (31.78\% and 45.80\%) remain substantially larger than plausible instrumentation noise. We therefore consider the comparative conclusions robust to reasonable levels of measurement uncertainty.

\textbf{Internal Validity.}
All configurations were executed under identical hardware, dataset, and software conditions with fixed random seeds. The paired design controls for stochastic variation across runs and strengthens causal interpretation within the studied environment. However, tightly controlled conditions may reduce natural variability compared to real-world deployments.

\textbf{External Validity.}
The evaluation was conducted using one CNN architecture, the CIFAR-10 dataset, and a single GPU type. While suitable for controlled experimentation, results may not directly generalize to larger architectures, different datasets, or alternative hardware environments. Replication across diverse ML workloads would strengthen generalizability.

\textbf{Conclusion Validity.}
Each configuration was executed ten times using a paired design. Although the sample size is modest, paired differences were consistent and yielded narrow confidence intervals, supporting the stability of the observed effects within this experimental setting.

\section{Conclusion}

This study presents initial empirical evidence that software-engineering practices may influence the environmental footprint of ML applications. We examined how two resource-leak smells in TensorFlow and Keras—Improper Model Reuse (IMR) and Unreleased Tensor References (UTR)—affect estimated electricity consumption and CO$_2$ emissions under controlled conditions.

Across the evaluated configurations, smell-injected models consistently consumed more electricity and produced more CO$_2$ than their clean counterparts. IMR increased electricity consumption by approximately 32\%, while UTR increased it by approximately 46\%, with proportional increases in emissions. 
While the study is limited to a single architecture, dataset, and hardware configuration, the findings suggest that resource-leak smells can introduce measurable sustainability overhead without functional benefit. These preliminary results motivate broader investigations across diverse ML workloads and infrastructure settings. Integrating resource-lifecycle management and sustainability-aware evaluation into ML development workflows may therefore represent a promising direction toward more environmentally responsible ML engineering.

Future work will extend this study along several dimensions. First, we plan to evaluate additional ML architectures (e.g., deeper CNNs and transformer-based models) and larger datasets to assess whether the observed sustainability overhead scales with model complexity. Second, we aim to systematically analyze a broader set of ML-specific code smells under controlled conditions to determine whether similar environmental patterns emerge. Finally, combining controlled experimentation with large-scale repository mining would enable quantifying the real-world prevalence and practical impact of resource-leak smells in production ML systems.

\bibliographystyle{ACM-Reference-Format}
\bibliography{sample-base}

\appendix
\section{Detailed Experimental Results}
\label{appendix:raw_results}

This appendix provides the full per-run measurements supporting the results in Section~\ref{sec:results}. Energy is reported in kWh and CO$_2$ in kg, consistent with CodeCarbon outputs.


\begin{table}[h]
\centering
\caption{Per-run energy consumption (kWh) and CO$_2$ emissions (kg).}
\label{tab:appendix_sust_raw}
\scriptsize
\setlength{\tabcolsep}{4pt}
\renewcommand{\arraystretch}{1.05}
\begin{tabular}{c ccc ccc}
\hline
\multirow{2}{*}{Run} & \multicolumn{3}{c}{Energy (kWh)} & \multicolumn{3}{c}{CO$_2$ (kg)} \\
 & Baseline & IMR & UTR & Baseline & IMR & UTR \\
\hline
1  & 0.025747 & 0.033276879 & 0.037246 & 0.011654 & 0.015062 & 0.016858224 \\
2  & 0.025519 & 0.033118433 & 0.037766 & 0.011550 & 0.014990 & 0.017093775 \\
3  & 0.025835 & 0.033778712 & 0.037331 & 0.011694 & 0.015289 & 0.016896926 \\
4  & 0.025589 & 0.034846279 & 0.037169 & 0.011582 & 0.015772 & 0.016823317 \\
5  & 0.025470 & 0.034870212 & 0.037932 & 0.011528 & 0.015783 & 0.017168825 \\
6  & 0.025671 & 0.033801634 & 0.037356 & 0.011619 & 0.015299 & 0.016908085 \\
7  & 0.025721 & 0.033505286 & 0.037629 & 0.011642 & 0.015165 & 0.017031792 \\
8  & 0.025692 & 0.033772977 & 0.037201 & 0.011629 & 0.015286 & 0.016838094 \\
9  & 0.025662 & 0.033113003 & 0.037019 & 0.011615 & 0.014988 & 0.016755532 \\
10 & 0.025558 & 0.033883028 & 0.037279 & 0.011568 & 0.015336 & 0.016834567 \\
\hline
\end{tabular}
\end{table}


\end{document}